\documentclass{emulateapj}
\usepackage{apjfonts}

\shortauthors{Helfand et al.}
\shorttitle{The MAGPIS Survey}

\slugcomment{Submitted to the Astronomical Journal}

\begin{document}
\title{MAGPIS: A Multi-Array Galactic Plane Imaging Survey}

\author{
David J. Helfand\altaffilmark{1},
Robert~H.~Becker\altaffilmark{2,3},
Richard~L.~White\altaffilmark{4},
Adam Fallon\altaffilmark{1}, \&
Sarah Tuttle\altaffilmark{1}}
\email{djh@astro.columbia.edu}

\altaffiltext{1}{Dept. of Astronomy, Columbia University, New York, NY 10027}
\altaffiltext{2}{Physics Dept., University of California, Davis, CA 95616}
\altaffiltext{3}{IGPP/Lawrence Livermore National Laboratory}
\altaffiltext{4}{Space Telescope Science Institute, Baltimore, MD 21218}

\begin{abstract}

We present the Multi-Array Galactic Plane Imaging Survey (MAGPIS), which maps
portions of the first Galactic quadrant with an angular resolution, sensitivity
and dynamic range that surpasses existing radio images of the Milky Way
by more than an order of magnitude. The source detection threshold at
20~cm is in the range 1-2~mJy over the 85\% of the survey region
($5\arcdeg<l<32\arcdeg, |b|<0\fdg8$) not covered by bright extended 
emission. We catalog over 3000 discrete sources (diameters mostly $<30\arcsec$) and present an atlas of $\sim 400$ diffuse emission
regions. New and archival data at 90~cm for the whole survey area are also
presented. Comparison of our catalogs and images with the MSX
mid-infrared data allow us to provide preliminary discrimination between
thermal and non-thermal sources. We identify 49 high-probability
supernova remnant candidates, increasing by a factor of seven the number
of known remnants with diameters smaller than $5\arcmin$ in the survey
region; several are pulsar wind nebula candidates and/or very small diameter 
remnants ($D<45\arcsec$). We report the tentative identification
of several hundred \ion{H}{2} regions based on a comparison with the mid-IR data;
they range in size from unresolved ultra-compact sources to large complexes
of diffuse emission on scales of half a degree. In several of the latter
regions, cospatial nonthermal emission illustrates the interplay
between stellar death and birth. We comment briefly on plans for followup
observations and our extension of the survey; when complemented by
data from ongoing X-ray and mid-IR observations, we expect MAGPIS to provide the
most complete census yet obtained of the birth and death of massive stars in 
the Milky Way.

\end{abstract}


\section{Introduction}

The Milky Way is a galaxy of stars radiating most of their energy at optical
wavelengths. But from stellar birth to stellar death, from the vast reaches of
interstellar space to the tiniest of stellar corpses, radio and X-ray
observations provide crucial diagnostics in our quest to understand the
structure and evolution of our Galaxy and its denizens. These two spectral
regimes are particularly crucial for studying massive stars: throughout their
lives, stellar Lyman continuum
photons produce \ion{H}{2} regions with their associated free-free radio emission,
while stellar wind shocks produce X-rays; in death, the remnants of
supernovae are the brightest radio and X-ray sources in the Galaxy.
Furthermore,
the Galaxy is largely transparent in the radio and hard X-ray bands, giving
us an unobstructed view through the plane, even at $b=0\arcdeg$. We
are in the process of conducting a large-scale survey of the Galactic plane
at X-ray wavelengths with XMM, the first results of which have been
reported elsewhere (Hands et al.\ 2004). Here we describe a complementary effort
to provide a new, high-resolution, high-sensitivity view of centimetric radio
emission in the Milky Way.

While significant progress has been made recently in surveying the 
extragalactic radio sky ({\it e.g.}, NVSS, SUMSS, and {\sl FIRST}), the 
Galactic plane still remains inadequately explored. Even though the NVSS 
(Condon et al.\ 1998) covered the plane, it did so in snapshot mode with $uv$ 
coverage insufficient to achieve high dynamic range (typical values achieved
are $\sim$30:1). The Canadian Galactic Plane Survey project (English et al.\ 
1998) is covering a large region of the plane in the second quadrant with 
better dynamic range, but with a resolution of only 
$65\arcsec$ and limited sensitivity in the continuum. The third
and fourth Galactic quadrants are being surveyed using Parkes and the Australia
Compact Telescope Array in the Southern Galactic Plane Survey 
(McClure-Griffiths at al. 2001), although the angular resolution is only 
$\sim 2\arcmin$ and the $5\sigma$ detection threshold is $\sim 35$~mJy in 
the continuum.

Fifteen years ago, we used observations originally taken by Dicke et al.\ 
(unpublished) in the B-configuration of the Very Large Array\footnote
{The VLA is a facility of the National Radio Astronomy Observatory which
is operated by Associated Universities, Inc. under cooperative agreement with
the National Science Foundation.}
(supplemented by additional 20~cm and 6~cm time awarded to us) to
produce a catalog of over 4000 compact sources within a degree of the plane in
the longitude range $-20\arcdeg<l<120\arcdeg$ (Becker et al.\ 1990; 
Zoonematkermani et al.\ 1990; White, Becker, and Helfand 1991; Helfand et
al. 1992). Although the original analysis provided maps that were complete 
only to $\sim 20$ mJy at 20 cm, this remains the 
highest resolution and most sensitive census of compact sources over a large 
segment of the Galaxy. Comparison with the IRAS survey led us to identify
more than 450 ultracompact \ion{H}{2} regions, over 100 new planetary nebulae (which
fill in the gap near $b=0$ caused by extinction in optical searches -- 
Kistiakowsky and Helfand 1995), and, along with 90 cm maps we obtained covering
a small portion of the longitude range, more than a dozen new supernova remnant
candidates.

Motivated by the torrent of new, high-resolution mid-infrared data from
the GLIMPSE Legacy survey with Spitzer (Benjamin et al.\ 2003) and taking 
advantage of modern data analysis algorithms developed for our FIRST survey 
(Becker, White \& Helfand 1995; White et al.\ 1997), we have recently completed a
reanalysis of all of the existing snapshot data (over 3000 individual pointings
including some new data designed to fill holes and improve quality in poorly
covered regions). This work yielded 6 and 20~cm catalogs with over 6000 entries
and flux density thresholds nearly a factor of two below those of the original 
analysis (White, Becker \& Helfand 2005). Nonetheless, the 
single-configuration, snapshot nature of these observations renders the data 
problematic for all but the most compact radio sources in the plane.

A high-sensitivity, high-resolution, high-dynamic-range map of the
radio continuum emission from the Galactic plane is now possible with a
relatively modest investment of telescope time owing to advances in the
VLA receivers over the last decade, the implementation of the highly
efficient ''survey mode" slewing algorithm, and improvements to the AIPS
software package. We have begun to make this possibility
a reality by producing a $5\arcsec$-resolution image of 27 degrees
of Galactic longitude in the first quadrant. Our plan over the coming several 
years is to extend this survey over the entire Spitzer GLIMPSE longitude range
in the north, covering $5\arcdeg<l<65\arcdeg$. This Multi-Array Galactic 
Plane Imaging Survey or MAGPIS (a moniker appropriate for the authors whose
careers have been based on collecting random shiny objects gathered from
overflights of much of the celestial sphere in several regions of the
electromagnetic spectrum) is designed
to provide a definitive archive of the Galactic sky at 20~cm.

In Section 2 we describe the survey parameters and the data acquired to date;
in addition, we discuss complementary datasets we have used in our
analysis and introduce the MAGPIS website, which offers comprehensive
access to all of our data products. Section 3 outlines our analysis strategy, 
presents the imaging results, and provides a statistcal characterization of the
survey sensitivity threshold and dynamic range. We then discuss our detection 
algorithms for both discrete and diffuse sources, and present the source 
catalogs as well as an atlas for all extended emission regions
(\S4). Section 5 includes a discussion of a preliminary comparison between
MAGPIS and the MSX mid-IR data, and previews the prospects for a more complete 
census of \ion{H}{2} regions in the first quadrant. This is followed by a discussion
(\S6) of the nonthermal emission regions detected in our survey, including the 
discovery of several dozen new supernova remnant candidates. We summarize our 
results in Section 7.

\section{The MAGPIS Survey: Design and Data Aquisition}

As noted above, radio emission is a prominent signature of massive stars;
\ion{H}{2} regions, pulsars, supernova remnants, and black hole binaries are
all the products of O and early B stars that have a small scale height.
This fact, coupled with constraints on the total observing time available, has
led us to restrict our Galactic latitude coverage to $|b|<0\fdg8$.
This is greater than the OB star scale height (Reed 2000) for
all distances beyond 3~kpc and covers a region up to $z=\pm 230$~pc at the 
solar circle on the far side of the Galaxy (we adopt $R_{\odot} = 8.5$~kpc
throughout).

\subsection{The 20~cm data}

Our first tranche of Galactic longitude, $32\arcdeg>l>19\arcdeg$, was
chosen to complement our first X-ray data set and to explore the tangent
to the Scutum spiral arm. The second segment we have completed covers the region
$19\arcdeg>l>5\arcdeg$; we stopped at $5\arcdeg$ mainly because the central
regions have been reasonably well-mapped previously.  We intend to continue
the survey as time becomes available, first to the GLIMPSE upper longitude 
limit of $l=65\arcdeg$ and later to both higher and lower longitudes.

Data are collected in the B-, C-, and D-configurations of the VLA operating
in pseudo-continuum mode at 20~cm; two 25-MHz bandwidths centered at
1365~MHz and 1435~MHz are broken into seven 3~MHz channels to minimize
bandwidth smearing as well as to reduce significantly our sensitivity to
interference. The loss of a factor of two in bandwidth over the standard 
continuum mode is not important, since virtually all maps are 
dynamic-range, rather than sensitivity, limited. 

\begin{figure*}
\epsscale{0.7}
\plotone{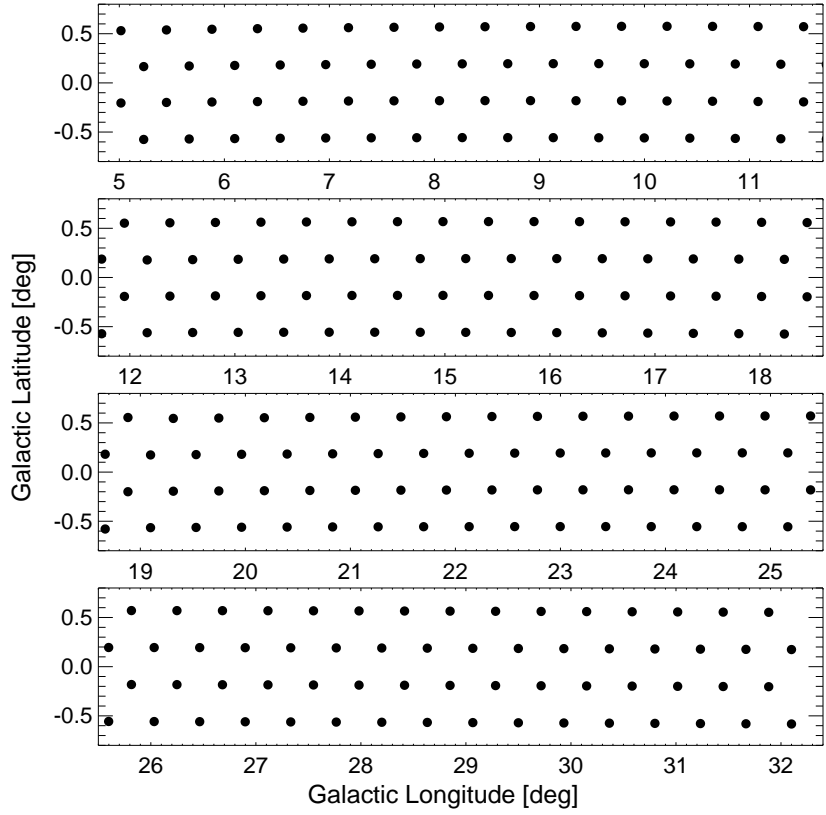}
\caption{
The hexagonal grid of 252 VLA pointing centers used for the MAGPIS 20~cm survey.
}
\label{fig-grid}
\end{figure*}

\begin{figure*}
\epsscale{0.7}
\plotone{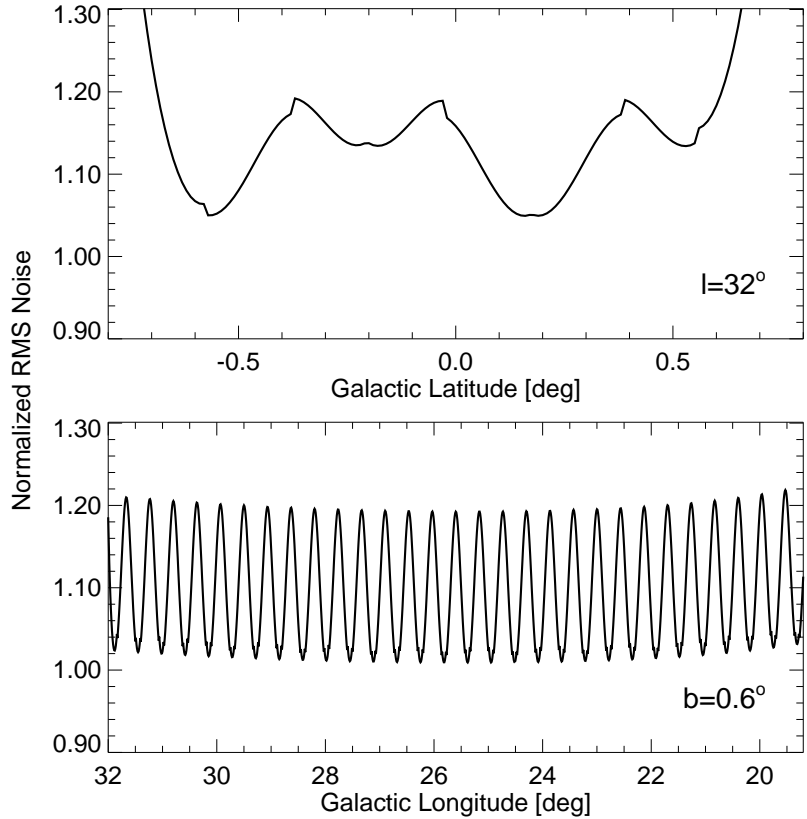}
\caption{
The variation in the rms noise as a function of
position after the overlapping images have been co-added.
The rms is normalized to unity at field center for a single
pointing.  The top panel shows a cut in latitude at the edge of the
survey ($l=32\arcdeg$), and the bottom panel shows the rms along a
line passing near the field centers at $b=0\fdg6$.  The
rms noise is uniform, with a peak-to-peak variation of only
$\sim \pm10$\% except at the edges of the surveyed area.
}
\label{fig-coverage}
\end{figure*}

The pointing pattern is displayed in Figure~\ref{fig-grid}. The close-packed hexagonal array
provides uniform coverage with a peak-to-minimum variation in sensitivity of
$<20\%$  (after co-adding of adjacent images ---
see Fig.~\ref{fig-coverage}). We observe each location four times in each of the
three configurations spaced roughly equally in hour angle over a range $\pm4$ 
hrs to maximize $uv$ coverage; the result is an average of $\sim12$ minutes 
per field per configuration, providing a theoretical noise level ($\sim 0.08$ 
mJy) far below the dynamic range limit of the maps. In the second round of 
observations we have saved observing time by using the full 12 minutes per 
field in the B configuration, but reducing the integration time by a factor of 
two in the two lower-resolution configurations (while maintaining the 
observational cadence at multiple hour angles). This reduces our sensitivity 
by $\sim 20\%$ in the least-populated map regions, although, again, most of 
the images are dynamic-range limited and the effect on the final source 
catalog is minimal.

A total of 165 hours of time has been accumulated to date in the MAGPIS
project. Table~1 lists the observing epochs and configurations used to
construct the 252 individual images, which cover an area of over 42~deg$^2$.

\tabletypesize{\scriptsize}
\begin{deluxetable}{ccccc}
\tablecolumns{5}
\tablecaption{Observing Log}
\tablehead{
& \multicolumn{4}{c}{VLA Configuration} \\
\colhead{Description} & \colhead{B} & \colhead{C} & \colhead{D} & \colhead{BnC}
}
\startdata   
Phase 1, 20~cm\tablenotemark{a}
               & Mar--Apr 2001 & Aug--Sept 2001 & Aug--Sept 2000 & May 2001\\
               &   31 hrs      &   28 hrs       &   28 hrs       &   1.5 hrs\\
Phase 2, 20~cm\tablenotemark{b}
               & Jan 2004      & Feb--Mar 2004  & Apr 2003       & \\
               &   36 hrs      &   19 hrs       &   18 hrs       & \\
Phase 1, 90~cm\tablenotemark{c}
               &               & Sept 2001      & & \\
               &               &   3.5 hrs      & & \\
\enddata   
\tablenotetext{a}{20 cm Phase 1: $19\arcdeg<l<32\arcdeg \, , \, |b|<0\fdg8$}
\tablenotetext{b}{20 cm Phase 2: $5\arcdeg<l<19\arcdeg \, , \, |b|<0\fdg8$}
\tablenotetext{c}{90 cm Phase 1: $20\arcdeg<l<33\arcdeg \, , \, |b|<2\arcdeg$}
\end{deluxetable}
\tabletypesize{\footnotesize}

\subsection{The 90~cm data}

Even with high-quality images, a single frequency is insufficient to
identify source classes unambiguously and to disentangle thermal and nonthermal
emission in crowded regions. As part of our initial observation program for
MAGPIS, we obtained 3.5 hours of 90~cm pseudo-continuum observations in the C 
configuration of the VLA during September of 2001. Eight pointings were used
to cover the region $20\arcdeg<l<33\arcdeg, |b|<2\arcdeg$. The data were
reduced using a $15\arcsec$ pixel size and have a resolution of
$\sim 70\arcsec$. 
In addition, we retrieved from the VLA archive 90~cm data originally taken
by Brogan et al.\ (2005) that cover the remainder of our current survey area
($3\fdg6<l<20\arcdeg, |b|<2\arcdeg$). These data were reduced using a 
$6\arcsec$ pixel size and have a resolution of $\sim 25\arcsec$.

\subsection{The mid-IR data}

\begin{figure*}
\epsscale{0.9}
\plotone{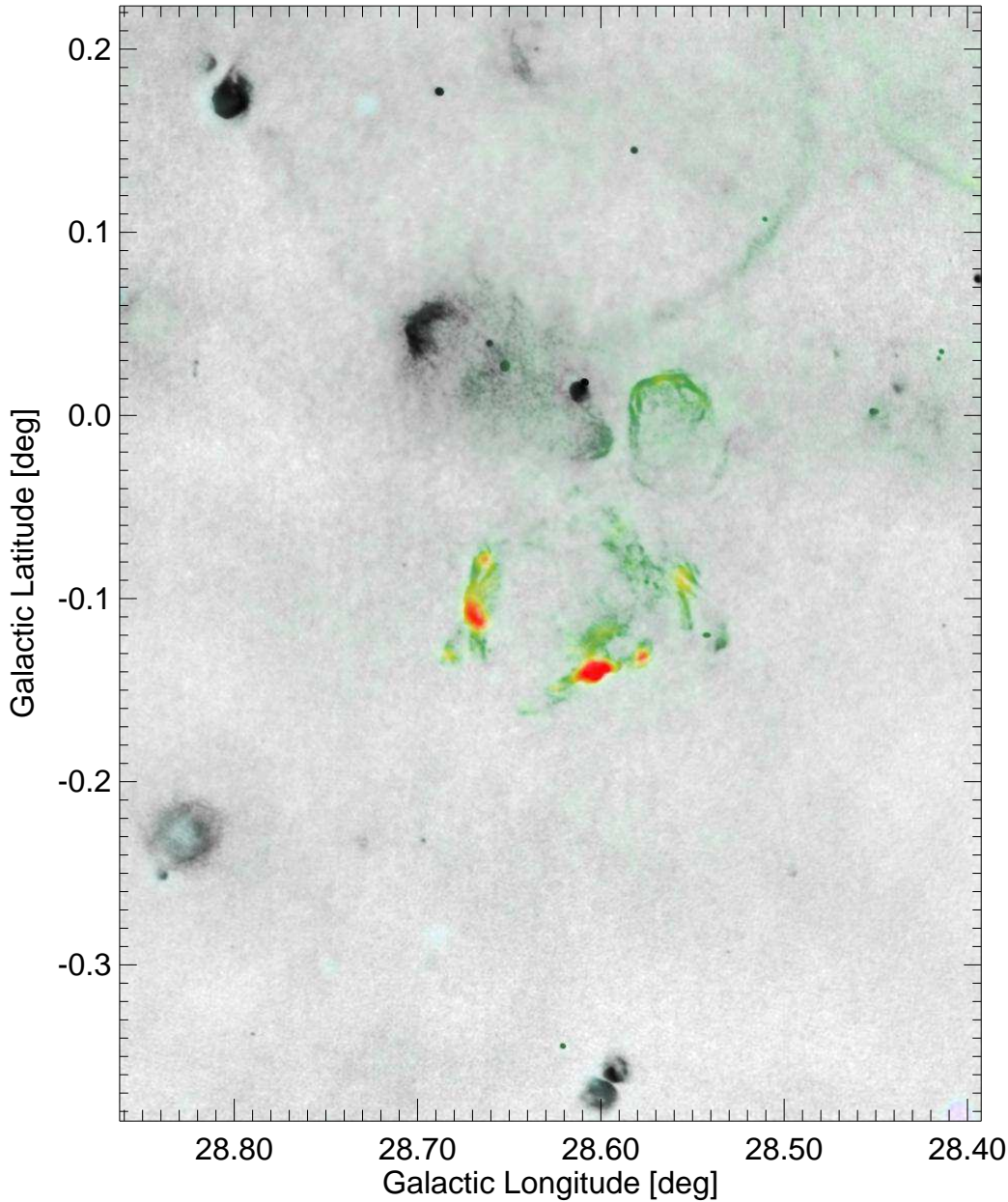}
\caption{
Combined radio-IR image demonstrating the separation of thermal and
non-thermal emission.  The radio image is used to set the intensity
of the displayed image, while the radio-IR flux ratio is used to
set the color hue and saturation.  Objects with strong IR emission
(typical of thermal radio sources) appear in black and white, while
objects that have absent or weak IR emission (typical of nonthermal
radio sources) appear in colors ranging from green to red depending
on the upper limit on the radio-to-IR ratio.  Both the known SNR
G28.6-0.1 and a previously undiscovered remnant at G28.56-0.01 are
apparent, as are a half-dozen thermal sources with varying morphologies.
}
\label{fig-msxcolor}
\end{figure*}

We have retrieved the mid-IR images and catalogs of the Midcourse Space 
Experiment (MSX -- Price et al.\ 2001) from the IPAC database for the regions
our survey covers to date. For ease of comparison, we have regridded the E-band
($20\,\mu$m) data onto the same $l,b$ grid used to present the primary MAGPIS 
images. We have also constructed ratio maps for the 20~cm and $20\,\mu$m data for
use in separating thermal and nonthermal emission. An example of such an image
is displayed in Figure~\ref{fig-msxcolor}. High values of the 
radio-to-IR ratio generally indicate nonthermal radio emission such as is
produced by supernova remnants, while low values tend to highlight dusty \ion{H}{2} 
regions, although pulsar wind nebulae, dusty old supernova remnant shells, and
dust-free \ion{H}{2} regions can in principle exhibit intermediate ratios. We defer
a quantitative discussion of the comparison of the radio and mid-IR emission
to a future paper.

\subsection{The MAGPIS website}

Consistent with our past practice, the raw VLA data on which MAGPIS is based
have been available in the VLA archive from the day they were taken. To
facilitate use of these data by the broadest possible community, we have
constructed the MAGPIS website (\url{http://third.ucllnl.org/gps}), which presents
our data products in easily accessible forms. In addition to the 
full-resolution 20~cm images, the site provides the complementary 90~cm images,
the regridded MSX $20\,\mu$m images, and an image atlas of diffuse emission
regions (see below). The single-configuration 6 and 20~cm images from
our earlier snapshot surveys (White, Becker \& Helfand 2005)
are also available.  Images can be displayed with 
user-specified coordinates, box sizes, and intensity scales or can be downloaded
as FITS files. The full discrete-source and diffuse catalogs are available for 
retrieval or through a search query function, as are our catalogs and 
publications from our earlier snapshot survey work. We expect to add our 
XMM X-ray survey data and the Spitzer GLIMPSE survey images and catalogs as 
they become available.

\section{The MAGPIS Images}

In contrast to the extragalactic radio sky, which is rather sparsely populated 
by mostly compact sources, radio emission in the Galactic plane is
dominated by bright, diffuse \ion{H}{2} regions and supernova
remnants. Thus, the single-snapshot observations and two-dimensional mapping
approximations that worked well in the FIRST and NVSS surveys are inadequate
for producing high-dynamic-range images for MAGPIS. In this case, the VLA data 
must be treated as a three-dimensional data set. In practice, 3-d distortions
scale with offset from the image center; thus, one way to minimize 3-d effects
is to tile the VLA's $30\arcmin$ primary beam with many small images. We have
used a grid of 21 by 21 images, each of which is 128 by 128 pixels in size. 
Our initial data were reduced on a Sun Ultra 60, with each image requiring
$\sim12$ hours to CLEAN. We subsequently migrated the analysis to a 
dual-processor Pentium 4 computer that is approximately seven times
faster. Since the images are greatly improved by self-calibration, each
field has to be reprocessed several times. 

Even with data from the D configuration, the resulting maps suffer
missing flux from large-scale structure ($\gg 1\arcmin$) to which the VLA is 
insensitive. To correct for this deficiency, we combined the VLA images with 
images from a 1400 MHz survey made with the Effelsberg 100-m telescope 
($\sim 7\arcmin$ angular resolution). The AIPS task IMERG makes FFTs of both 
the VLA and Effelsberg images, combines the derived FFT amplitudes, and then 
converts back to the image plane to produce the final individual images. We use
a $6\farcs2 \times 5\farcs4$ restoring beam on maps with a
pixel size of $2\farcs$. The individual maps are ultimately summed 
and rebinned to produce mosaic images in Galactic coordinates.  The dynamic
range varies somewhat with location but, measured as a ratio of the peak
flux in the brightest source to the full image rms, is typically in excess
of 1000:1 in a 1 deg$^2$ image. Over most of the images, 1~mJy point
sources are easily detected.

\section{The MAGPIS Catalogs}

The large, diffuse emission features and variable background, coupled with 
source size scales ranging from arcseconds to degrees, render impractical the
type of automated source detection algorithms applied to extragalactic
radio surveys. Thus, we have employed the human eye-brain detection
system to search the 16.7 million MAGPIS beam areas for radio sources.
We divided the problem into two parts: the detection and cataloging of
discrete objects less than a few beam areas in size and unconfused by
extensive diffuse emission, and regions of sky in which significant diffuse
emission is present.

\subsection{Discrete source detection}

A square field was defined around each 
candidate discrete source.  In cases where it was impossible to isolate a 
single emission peak (e.g., for overlapping or closely clustered sources), 
multiple sources were included in one field and the field was flagged as 
"multiple''. The default field size was $34\arcsec\times34\arcsec$,
but this size was adjusted for larger sources (increased), for high 
density areas (decreased), or for other reasons (increased or decreased on a 
case by case basis) such as nearby bad pixels, proximity to the edge of the 
maps, etc. For the entire survey area this process yielded 2628 single-source 
fields, and 467 multiple-source fields.

The AIPS task HAPPY (see White et al.\ 1997) was then run on each of the
fields.  In HAPPY, a local rms level was calculated for each field
using an area of three times the input field area; a minimum detection 
level of five times the local rms was set for each field.

Using the HAPPY output, we rejected any source with a fitted peak flux, $F_p < 
1.0$~mJy or $\le 5.0$ times the local rms, whichever is higher. We also rejected
any source with a fitted minor axis less than $3\farcs5$ 
(the beam minor axis is $5\farcs2$, and experience from 
the use of HAPPY in the FIRST survey shows that the vast majority of such 
``skinny'' sources are spurious sidelobes); this only eliminated one source
that passed the $F_p$ and rms criteria. This process yielded a catalog of 3229 
sources.

Although restricting HAPPY to predetermined fields around candidate sources
should reduce the number of spurious detections, this method is still
susceptible to poor fits resulting from complex, extended emission as well as
areas of patterned noise near bright sources. To assess these potential causes 
of contamination, we flagged for further examination HAPPY solutions

\begin{enumerate}

\item when HAPPY fit more than one source in a single-source input field (67
fields, 144 sources);

\item when HAPPY fit more than two sources in a multiple-source input field 
(69 fields, 219 sources); or

\item when any fit not meeting (1) or (2) had a major axis 
$>15\arcsec$, a major to minor axis ratio greater than 2.0, 
or $F_{int}/F_p > 5.0$ (136 fields, 143 sources).

\end{enumerate}

In total, 506 fields were flagged and examined. Of these 338 were
determined to be good fits, 88 to be ``acceptable''  fits, 56 to be
artifacts or noise, and 24 to be extended rather than discrete
sources; the latter were moved to the extended source atlas (see below)
and the artifacts were deleted.

The 88 ``acceptable'' fits are all (in our best judgement) real radio
sources, but they are distinguished from good fits in that, upon inspection,
it is clear that the two-dimensional Gaussian employed by HAPPY is a poor
representation of the source surface brightness distribution. We report
their HAPPY-derived parameters in Table~2 for consistency, but flag
them accordingly.

The final catalog, presented in Table~2, includes 3149 discrete
sources.  A Galactic longitude $\pm$ latitude-based name (col.~1)
is followed by the peak and integrated flux densities from the
Gaussian fit (cols.~2 and~3), with the $S_i$ value flagged for the
``acceptable'' fits described above, and the estimated rms noise
level (col.~4).  The major and minor axes (full-width at half-maximum)
and position angle for the elliptical Gaussian complete the
morphological description of these compact sources.  The last two
columns give the infrared 8~$\mu$m and 21~$\mu$m flux densities for
sources with MSX matches (described further below in
\S\ref{section-msxmatch}).

\begin{figure*}
\epsscale{0.7}
\plotone{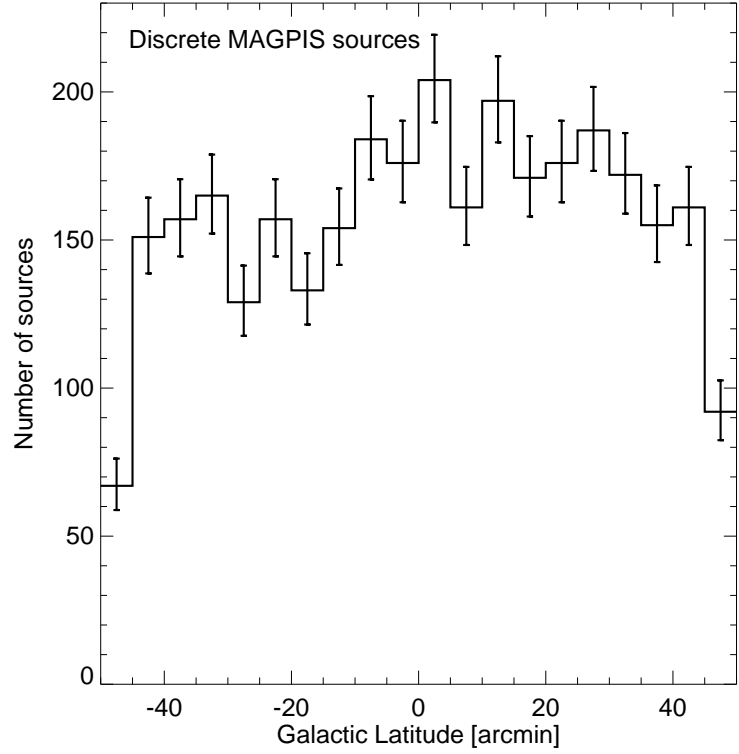}
\caption{
Galactic latitude distribution for the discrete source catalog.
Even in the Galactic plane the 20~cm radio sky is dominated by
extragalactic sources, so no strong latitude dependence is seen.
The counts fall off in the $|b|>45$~arcmin bins due to the drop in
sensitivity at the edge of the survey.
}
\label{fig-lathist}
\end{figure*}

Owing to the variable background and numerous regions of bright diffuse
emission, the threshold for discrete source detection varies significantly over
the survey region. We can obtain a mean value for the threshold by comparing
the source surface density with that of the FIRST survey. That survey covers
9033 deg$^2$ of the extragalactic sky and includes 781,450 sources not flagged 
as sidelobes, for a mean surface density of 86.54 deg$^{-2}$ at a flux density
threshold of 1.0~mJy\footnote{The snapshot images of the FIRST survey require 
the addition of a 'CLEAN bias' of 0.25~mJy to the measured flux densities. The 
greater $uv$ coverage achieved in the multi-array, multi-snapshot MAGPIS survey
should significantly reduce CLEAN bias, although it is improbable that the bias
is zero. Since, however, absolute calibration is unlikely to be accurate to 
better than 10\% in light of our addition of single-dish data, and none of our 
scientific projects require flux densities this accurate, we ignore CLEAN 
bias in this work.}. The mean source surface density of discrete sources in
MAGPIS is 74.8 deg$^{-2}$ considering the entire survey area of 42.1 deg$^2$
and 73.0 deg$^{-2}$ in the 35.6 deg$^2$ lying outside regions of diffuse
emission; the former value is higher owing to source clustering.
Matching this surface density while allowing for
several hundred true Galactic sources outside regions of diffuse emission
(see \S5.1), we find an effective discrete source threshold of $\sim 1.5$~mJy
that yields 58.6 extragalactic sources deg$^{-2}$ in {\sl FIRST}. Thus, our
survey is significantly incomplete between the minimum reported flux density
of 1~mJy  and $\sim 2$~mJy, but, over the 85\% of the area outside regions
of diffuse emission, it is largely complete above this range. Note that a
large majority of the discrete radio sources detected even within
$1\arcdeg$ of the Galactic plane are extragalactic objects; this is evident 
from the lack of a strong Galactic latitude dependence of our source counts seen
in Figure~\ref{fig-lathist}. Observations at other wavelengths are required to identify the
Galactic components of the discrete source population.

\subsection{The diffuse source atlas}

The elliptical Gaussians used in fitting the discrete sources are a poor
approximation to the surface brightness distributions for nearly all of
the more extended radio sources detected in our survey. Furthermore, for
sources extended by more than $\sim 60\arcsec$, our VLA $uv$ coverage
is inadequate to derive accurate flux densities, and the addition of the single
dish data, while an asset in making images, has unquantifiable effects on
derived flux densities. Thus, we again turn to the eye-brain system for
identifying diffuse sources and source complexes, and do not attempt to derive
accurate flux density measurements for these sources.

The entire survey region was examined by eye, and regions of extended emission
were identified and enclosed in square boxes ranging in size from
1 arcmin$^2$ to $48\arcmin \times 48\arcmin$. In some instances
regions are defined by a single coherent source, while in others a complex of
diffuse emission regions is included. A total of 398 such regions covering 
7.6 deg$^2$ were so identified. For each region, the
peak flux density, minimum flux density (a proxy for the noise level in
the region), total area, and net flux density were recorded; we emphasize that
these flux densities are not necessarily accurate reflections of integrated
source intensity and, in some regions, include the flux density of several
related -- or possibly unrelated -- sources; they provide only a rough
guide to source intensities. We have subtracted from the integrated flux 
density in each region the sum of the flux densities of the discrete sources 
from Table~2 that fall within the region; a table listing the cataloged
discrete sources within each region is available at the MAGPIS website.

\begin{figure*}
\epsscale{0.7}
\plotone{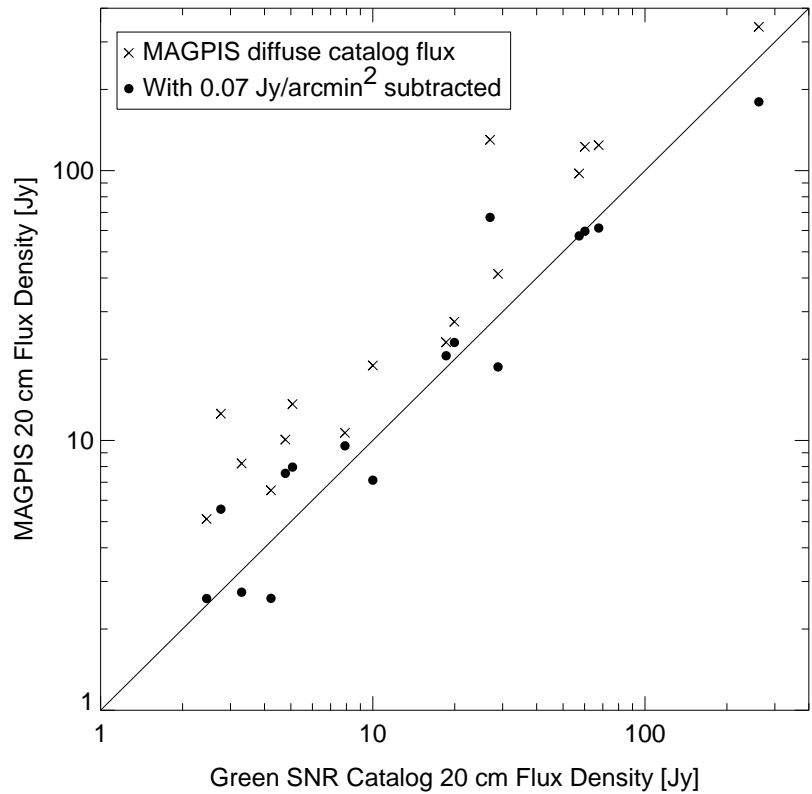}
\caption{
Flux densities from our diffuse source catalog for supernova remnants
from the Green (2004) catalog.  The x symbols show that our catalog
flux densities for these extended sources are typically higher than
the Green values by factors of two or more.  Subtracting a corrective
offset of 0.07~Jy/arcmin$^2$ from our flux densities improves the
agreement (dots).
}
\label{fig-snrflux}
\end{figure*}

In order to estimate the accuracy of our diffuse flux density estimates, we
have compared our flux densities for the 25 known supernova remnants in our
survey region with
those tabulated in Green (2004). We exclude remnants for which the
tabulated value is uncertain (listed with a ``?'' in Green's catalog), as well
as those that do not fall completely within our survey coverage. We scale
the 1.0~GHz flux densities listed in Green's catalog to our observing 
frequency of 1.4~GHz using the tabulated spectral indices. We find a
good correlation between the flux densities, albeit with an offset
that depends on the size of the remnant (Fig.~\ref{fig-snrflux}).
We conclude that the integrated flux densities listed for the diffuse
sources typically overestimate the true fluxes of very large sources
by factors of two or more due to backgrounds and confusing sources and
recommend caution when using them.

The diffuse regions are cataloged in Table~3.  A Galactic longitude$\pm$
latitude-based name is found in column 1. 
The box size in column
2 and an intensity scaling factor for display purposes (col.~3) precede
the brightest pixel value (col.~4) and its location (cols.~5 \& 6), the minimum
flux density recorded in the box (col.~7), and the integrated flux density
inside the box (col.~8). Column 9 provides names for known
supernova remnants.

Cleaving to the maxim that quantifies the relative information
content of words and pictures, we have constructed a diffuse source
atlas to accompany the full survey images on the MAGPIS website.
Here, Table~3 is reproduced with active links that allow the user
to overlay circles representing sources from the discrete source
catalog and contours of the $20\,\mu$m images from the MSX catalog
(Eagen et al.\ 2003).  Each image can also be downloaded as a FITS
file.

The website also includes large area ($4\fdg5\times1\fdg6$) JPEG
versions of the MAGPIS images with the diffuse region boxes overplotted.
It is difficult to display these high-dynamic-range images
with a single constrast stretch, and indeed the discrete sources are
almost invisible in these images, but they are nonetheless useful
for viewing the environment of the diffuse sources.

The MAGPIS discrete source catalog and diffuse source atlas provide
an improvement of more than an order of magnitude in both angular
resolution and sensitivity over existing Galactic plane survey data.
When combined with existing catalogs at other wavelengths along
with data from X-ray and infrared surveys currently underway, MAGPIS
will provide a resource for studying both thermal and nonthermal
processes that mark the evolution of massive stars in the Milky
Way. We have a number of followup projects underway; below we briefly
comment on the impact the survey is likely to have on our knowledge
of the \ion{H}{2} region and supernova remnant populations of the
Galaxy.

\section{Galactic Thermal Emission Regions: MAGPIS and Mid-IR Images}

The critical dependence of an \ion{H}{2} region's radio luminosity on the ionizing 
flux of its exciting star(s) allow for the contruction of a particularly pure 
census of massive star formation: the 20cm radio flux density 
falls by a factor of 300 between exciting star types O9.5 and B1 such that, at 
20 kpc, O-star \ion{H}{2} regions fall a factor of $>30$ above our survey threshold, 
while less-massive star-forming complexes (which produce at most B stars) fall 
a factor $>10$ below it\footnote{These numbers are for optically thin nebulae.
Optically thick \ion{H}{2} regions are self-absorbed at 20~cm such that only stars
earlier than O7 would fall above our threshold at the far side of the Galaxy.
Our old 6~cm snapshot survey allows us to find these sources down to spectral
type O9.5; see Giveon et al.\ (2005) for details.}. To separate the \ion{H}{2} regions
from the more numerous extragalactic source populations and the extended
regions of Galactic nonthermal emission requires observations at another 
wavelength. Our 6~cm snapshot survey is useful for the most compact sources
($D<15\arcsec$) but resolves out flux on larger scales. Since most
\ion{H}{2} regions also contain dust that is heated by the stellar radiation, the
mid-IR also can serve as a useful discrimnant.

\begin{figure*}
\epsscale{1.0}
\plotone{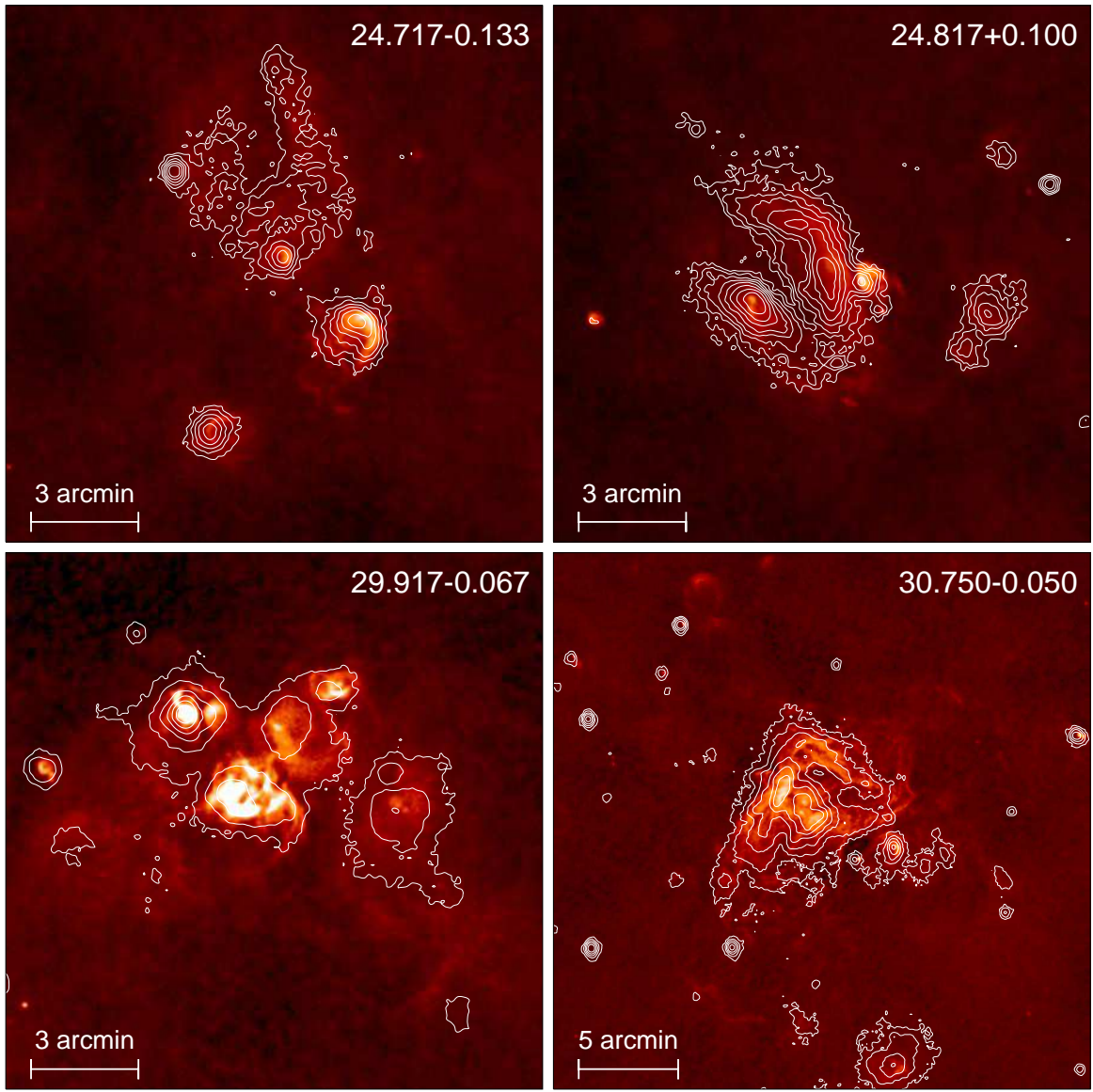}
\caption{
Examples of \ion{H}{2} complexes from our radio survey with the MSX $20\,\mu$m
image contours overlaid, showing the excellent radio-IR correspondence for
these thermal sources.
}
\label{fig-radioir}
\end{figure*}

Figure~\ref{fig-radioir} shows several examples of \ion{H}{2}
complexes from our radio survey with contours from the MSX $20\,\mu$m
images overlaid. The degree of correspondence is remarkably good
and provides a straightforward method for separating thermal and
nonthermal emission in star formation regions.  On the MAGPIS website
we also provide large-scale radio images with boxes marking the
previously published \ion{H}{2} regions collected in the Paladini
et al.\ (2003) meta-catalog.  It is clear that the MAGPIS data
(along with other radio and IR surveys) will enable the construction
of a vastly improved \ion{H}{2} region catalog.

We defer a detailed analysis of the \ion{H}{2} region population to a future paper;
here we provide some simple statistics for compact and ultracompact \ion{H}{2}
regions by matching our discrete source data to the MSX catalogs as an
indication of the wealth of information such a comparison contains. The
higher-resolution and greater sensitivity of the Spitzer GLIMPSE data
soon to become available will fill in the 2--8$\,\mu$m band and provide
crucial information in the most crowded regions.

\subsection{Match to the MSX $20\,\mu$m catalog}
\label{section-msxmatch}

The 20cm survey region is completely covered by the MSXPSCv2.3 "MSX6C"
(Egan et al 2003) data set.  We searched for MSX6C sources using a
search radius of $12\arcsec$ around each of the discrete 20~cm sources.
To be accepted as a match, the MSX6C source was required to have a quality 
flag of 2 in at least one of the four bands (see Lumsden et al.\ 2002).  If 
more than one MSX6C source fell within the search radius for a single 20~cm 
source, the MSX6C source closest to the 20~cm source was kept (this
only occurred once). A total of 376 MSX6C sources corresponding to 418
20~cm sources were matched in this manner.

To estimate the number of false matches we repeated the matching process using 
fake catalogs produced by shifting the MSX6C catalog $\pm10\arcmin$ and 
$\pm20\arcmin$ in longitude. Since, for example, the vast majority (78\%) of 
the MSX sources are stars detected only in the $8\,\mu$m band (very few of which
have radio counterparts), we can greatly reduce the false match rate by 
assessing the false rates separately for sources detected in different band 
combinations. We have followed the methology descibed in Giveon et al.\ (2004; 
see also White et al.\ 1991), to arrive at a false-match reliability criterion 
for each of the band combinations in which a 20~cm-MSX6C match existed. 
Using a reliability of $R>90\%$\footnote{This eliminates MSX sources detected
only in the $8\,\mu$m band as well as those detected in $8\,\mu$m and $12\,\mu$m 
only, and $8\,\mu$m, $12\,\mu$m, and $14\,\mu$m only. 
This removes 131 sources (67\% of which are false matches),
leading to a catalog of matches 
that is $>95\%$ reliable and $\sim 90\%$ complete.} we find 245
MSX6C sources (of which $\sim 8$ should be false) matched 
to 278 20~cm sources. Of these, 217 are single 20~cm-MSX6C matches, 23 are 
cases in which one MSX6C source matches two 20~cm sources, and 5 represent 
one MSX6C source matching three 20~cm sources.

\begin{figure*}
\epsscale{1.0}
\plotone{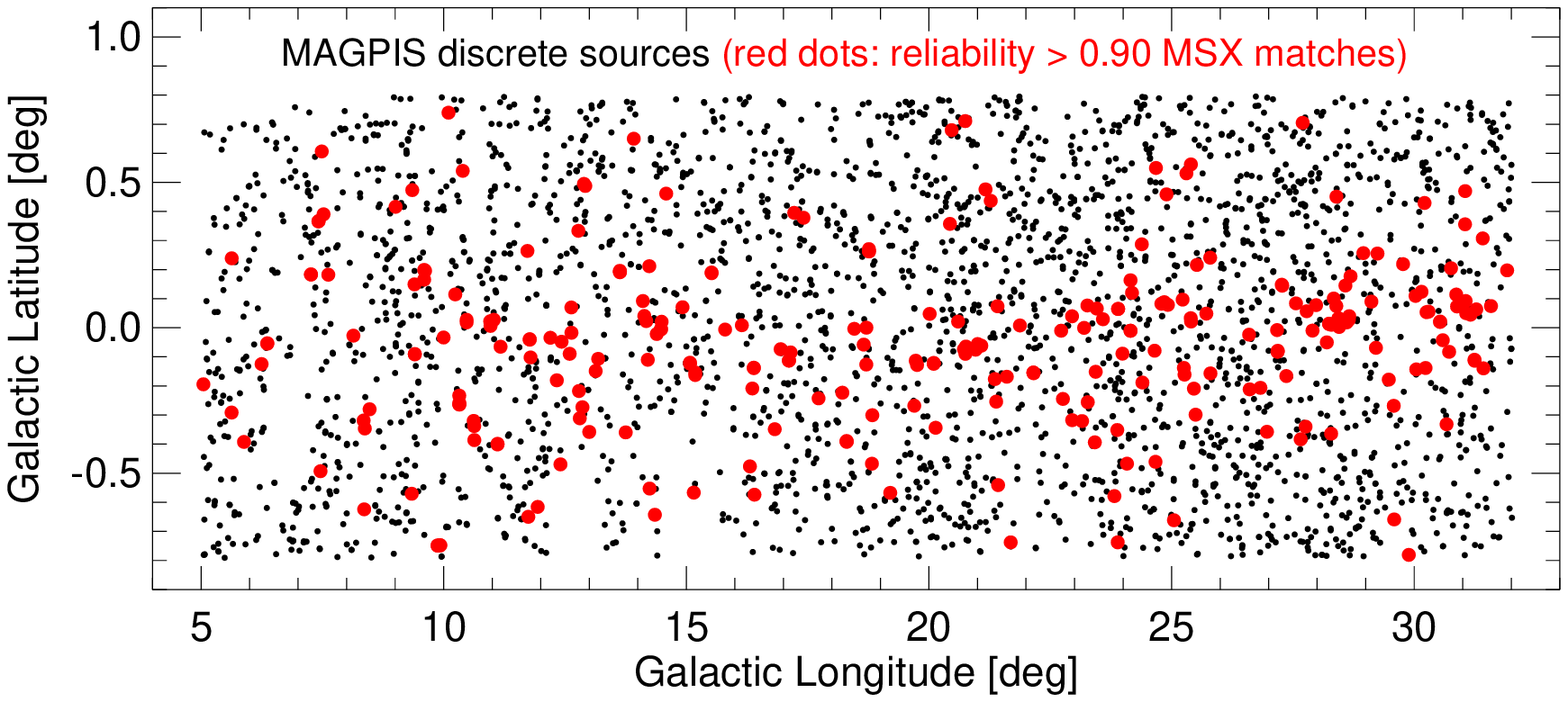}
\caption{
Sky distribution of MAGPIS 20~cm radio sources from the discrete
source catalog. Red dots show sources with confident (reliability
$>0.90$) infrared counterparts in the MSX catalog.  The radio-MSX
matches are clearly concentrated toward the plane.
}
\label{fig-skydist}
\end{figure*}

\begin{figure*}
\epsscale{0.7}
\plotone{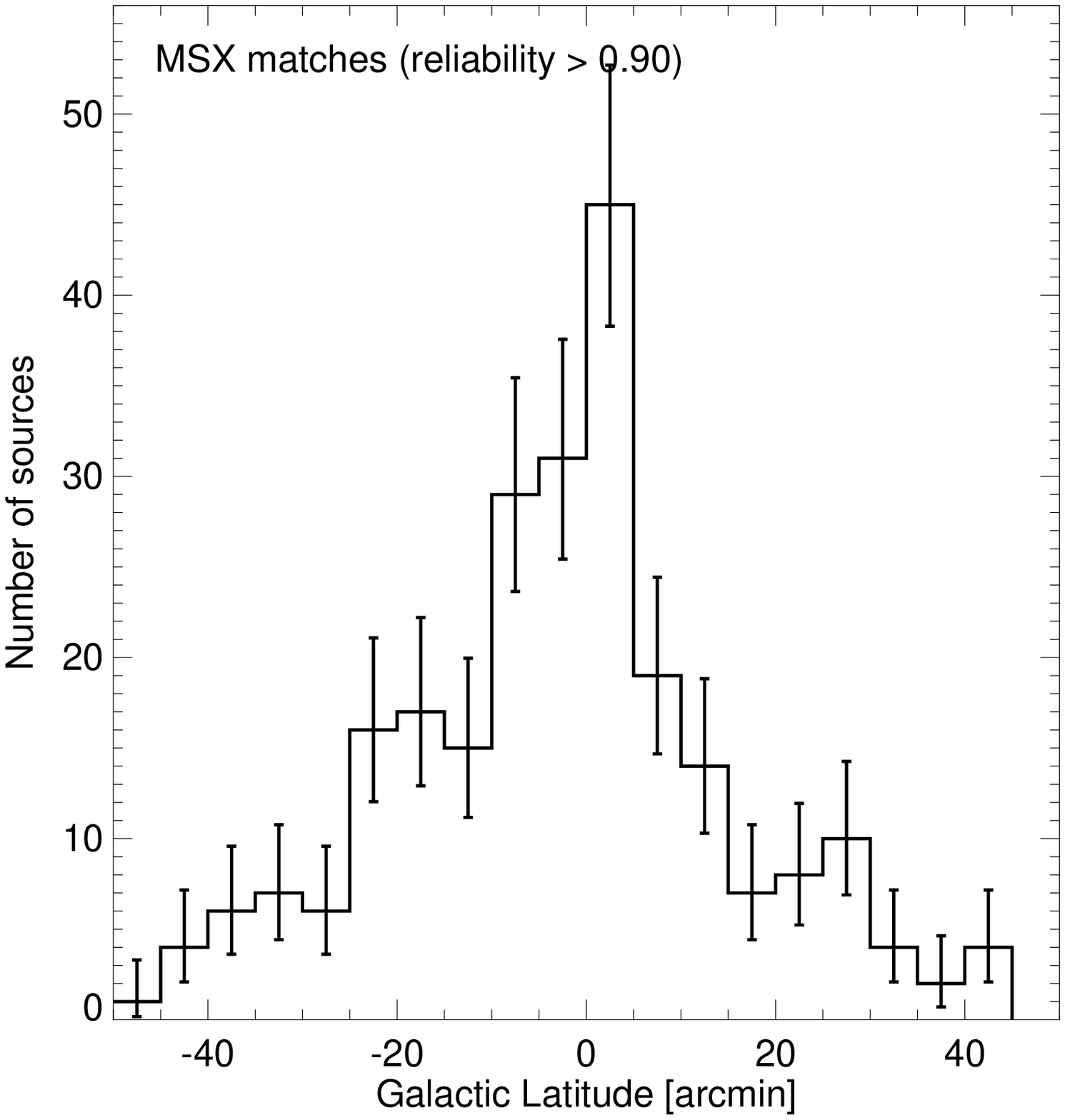}
\caption{
Galactic latitude distribution for the 245 MSX sources having reliable radio
matches.  These sources are highly concentrated toward the plane
of the Galactic disk.
}
\label{fig-msxlathist}
\end{figure*}

The distribution of the matched and unmatched sources on the sky is displayed
in Figure~\ref{fig-skydist}. While sources with infrared matches are found throughout the
latitude coverage, it is clear that they concentrate toward the plane.
The latitude distribution is displayed in Figure~\ref{fig-msxlathist}. The distribution peaks
at $b=0\arcdeg$ with a surface density of 22 sources deg$^{-2}$ (when regions obscured
by bright diffuse sources are excluded), and has a full width half maximumn of 
less than $15\arcmin$. Examination of the atlas of extended emission shows
that there are more thermal sources inside the 6.5 deg$^2$ subsumed by the
atlas images than outside, and most of these sources are not included in the
discrete source catalog. We estimate that there are a total of more than 600
distinct \ion{H}{2} regions in our 42 deg$^2$ survey area, although we defer to
a future publication the development of a detailed catalog and its analysis.

\section{Galactic Nonthermal Emission in MAGPIS}

Supernova remnants (SNRs) are among the brightest radio sources -- and the
brightest X-ray sources -- in the Galaxy. They are a dominant source of
mechanical energy input to the ISM, drive the Galaxy's chemical evolution,
and mark the birthsites of neutron stars and black holes. Yet our knowledge of
the Galactic population is woefully incomplete, owing to the low angular
resolution of previous radio and hard X-ray surveys of the plane, and the soft
spectral response of previous X-ray imaging observations. A total of 231
remnants appears in the latest catalog (Green 2004); Brogan et al.\ (2004) have
recently added three new remnants in one of our fields. The current rate of
discovery is a few remnants per year. However, based on 1) extragalactic SN rates
($\sim$ 1--2 per century) combined with SNR lifetimes (2.5-5$\times10^4$ yr), 
and 2) a detailed analysis of the current SNR distribution (Helfand et al.\ 
1989), we expect the total population to be between 500 and 1000. The youngest
remnant we know is 340 years old; four to seven younger ones exist somewhere
in the Galaxy.

MAGPIS can detect pulsar-driven remnants to a luminosity $10^{-4}$ that of the 
Crab Nebula at the edge of the Galaxy (or $\sim$ 10\% that of 3C58, the 
least luminous young Crab-like remnant known). For shell-like SNRs, our survey 
will be sensitive to all young remnants. For example, we
will detect remnants throughout the surveyed volume down to luminosities
0.01\% that of Cas A, and can even see a clone of the underluminous historical 
remnant SN 1006 at 20 kpc: it would appear as a $1\arcmin$-diameter source 
with a flux density of $\sim 25$~mJy. Our survey could detect
a remnant equivalent to SN87A from the time it was 3 years old anywhere in
the survey region, and would resolve such a remnant only 15 years after the 
explosion.

\begin{figure*}
\epsscale{0.9}
\plotone{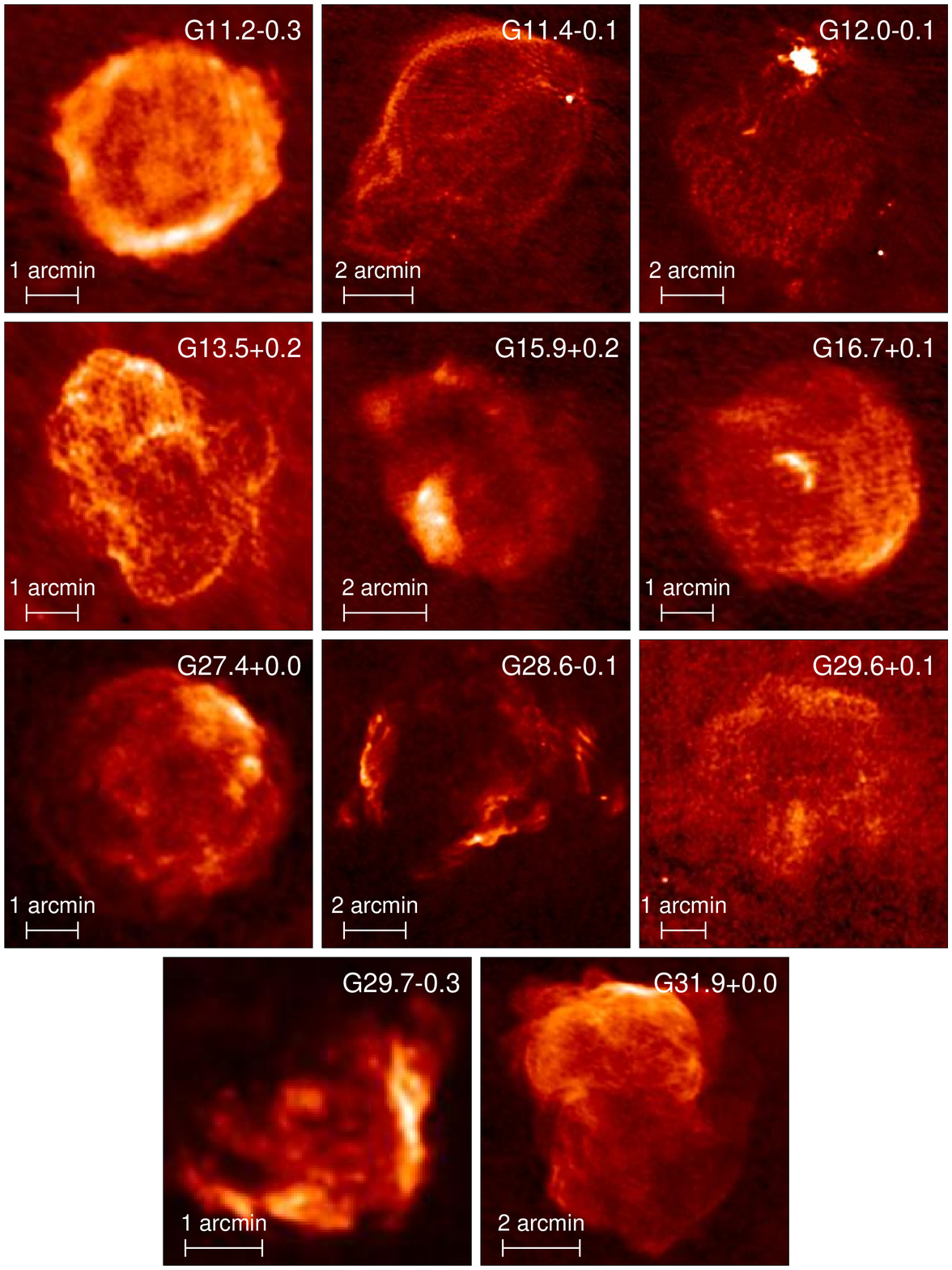}
\caption{
MAGPIS 20~cm images of the eleven known supernova remnants in the
survey area with diameters less than $10\arcmin$.
}
\label{fig-knownsnr}
\end{figure*}

Twenty-five known remnants fall within the current survey area, and all are
easily detected. The known remnants are indicated in Table~3; in many 
instances, the maps presented here are the best available. Images for the
eleven remnants smaller than $10\arcmin$ in diameter  --- most of
which lack 
high-resolution maps in the literature ---  are displayed in
Figure~\ref{fig-knownsnr}.

As can be seen by browsing the diffuse-source atlas, there are a large
number of shell-like sources detected in our survey. Without observations
at other wavelengths, however, it is impossible to separate the thermal
and non-thermal sources to derive a list of new SNR candidates. Fortunately,
as noted
above, we do have VLA data covering the entire region at 90~cm, as
well as the MSX mid-IR images. A simple qualitative comparison of these
three datasets (available for the reader at the MAGPIS website) allows us
to identify quickly high-probablity SNR candidates.

In Table~4, we present 49 new SNR candidates in our $27\arcdeg$ slice
of Galactic longitude. To derive this list, we have required:

\begin{itemize}

\item the object has a very high ratio of 20~cm to $20\,\mu$m flux (i.e., it
is typically undetectable in the 20$\,\mu$m MSX image);

\item the object has a counterpart in our 90~cm images with a similar
morphology and a higher peak flux density; and

\item the object has a distinctive SNR morphology. For shell-type remnants
we require at least half of a complete shell, while for the two pulsar
wind nebula candidates we see a centrally peaked brightness distribution.

\end{itemize}

For most of these candidates the data in columns 1, 3, 4 and 5 are
repeated from Table 3.  Column 2 gives the source diameter (as
opposed to the display box size in column 3, which is always larger).
Five of the the entries in this table are components of larger sources
listed in Table 3, with three associated with the large diffuse
complex at G19.60${-}$0.20 and two associated with G6.50${-}$0.48.

\begin{figure*}
\epsscale{0.9}
\plotone{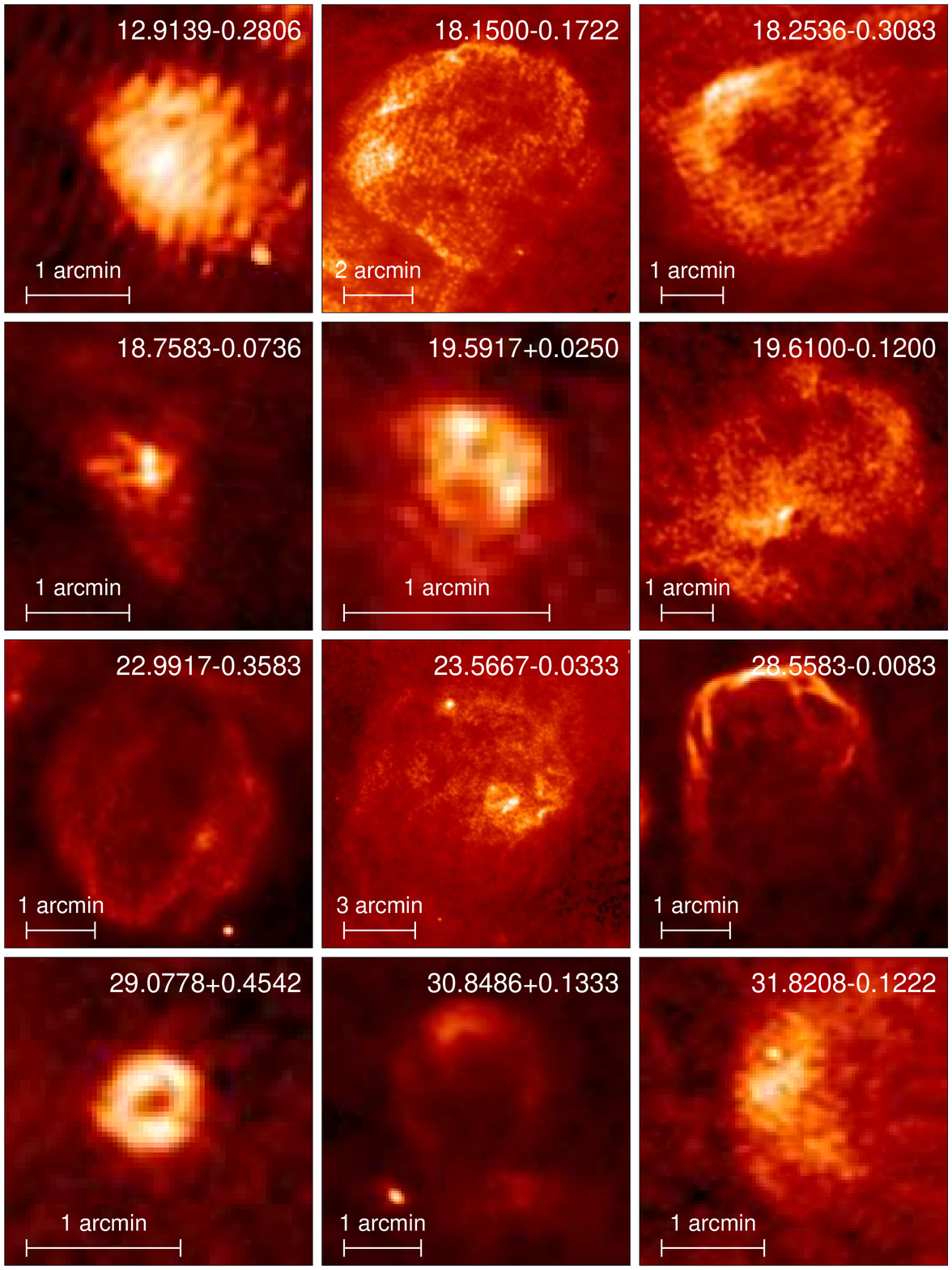}
\caption{
MAGPIS 20~cm images of 12 new supernova remnant candidates.
}
\label{fig-snrcand}
\end{figure*}

Images for a dozen candidates ranging in size from $40\arcsec$ to
$9\arcmin$ are displayed in Figure~\ref{fig-snrcand}. Not unexpectedly, the diameter
distribution for our remnant candidates varies markedly from that of the known
remnant population. Assuming that followup spectral and polarimetric
observations confirm the large majority of these sources as SNRs, we will
have tripled the number of known remnants in this region of the Galaxy.
However, while the number of remnants with $D \ge 10\arcmin$ will only
rise from 13 to 16, the number with $10\arcmin > D \ge 5\arcmin$ will
quadruple from 5 to 19, while the number with $D<5\arcmin$ will rise
more than sevenfold from 5 to 37.

Particularly interesting among these new SNR candidates are those that
may harbor young, high-$\dot E$ pulsars. In addition to the two PWN
candidates, there are two shell-like remnants with central diffuse
emission peaks highly reminiscent of composite SNRs. Given the core-collapse
SN rate in the Galaxy, we should expect to find $\sim 10$ neutron stars younger
than the Crab and 3C58 pulsars. While these new sources are significantly
dimmer than even the underluminous PWN 3C58, if they are at distances of
$\sim 15$~kpc, their luminosities are comparable. Also noteworthy
are the three shell-type remnants with diameters less than $1\arcmin$.
At 15~kpc, their diameters are $\sim 3$~pc, corresponding to an age of
$\sim 130$ years for a radio expansion rate comparable to that of SN87A.

The SNR candidates listed in Table~4 far from exhaust the nonthermal emission
features in our survey area; a roughly comparable number of filaments and
arcs with apparently nonthermal radio spectra and no IR counterparts are
seen. Furthermore, there are several regions
in which thermal and nonthermal features are cospatial; these will require
scaled-array observations at several frequencies to disentangle. Nontheless,
it is clear that high-dynamic-range, high-sensitivity observations
of the type reported here are essential for characterizing fully the Galactic
SNR population. 

\section{Summary and Future Prospects}

We have presented a centimetric image of the plane of the Milky Way in
the first quadrant that represents an improvement over existing
surveys by more than an order of magnitude in resolution, sensitivity, and
dynamic range. The survey detection threshold is 1 to 2~mJy over most of the
survey area. We identify over 3000 discrete radio sources and $\sim 400$ regions
of diffuse emission, presenting catalogs and atlases that quantify the
source properties. We include complementary 90~cm images over the entire 
survey region and provide a comparison with mid-IR data; taken together,
these latter two datasets help to separate thermal from nonthermal emssion 
regions. We find several hundred \ion{H}{2} regions in the survey area, many reported 
here for the first time. We also identify 49 high-probability supernova 
remnant candidates, including a seven-fold increase in the number of remnants with 
diameters smaller than $5\arcmin$ in the survey region. All of the survey's 
results are available at the MAGPIS website.

Considerable work remains to exploit fully the survey results. A complementary
hard X-ray survey over portions of this region is being conducted with 
XMM-Newton, and several followup obervations of interesting sources are
scheduled with XMM and Chandra. Scaled-array polarimetric and photometric 
observations with the VLA are required to confirm the SNR candidates. As the 
Spitzer GLIMPSE program images become available, further progress will be 
possible in identifying compact and ultra-compact \ion{H}{2} regions and in using
these to provide a census of the OB star population; higher frequency
observations with the VLA will be required to identify optically thick \ion{H}{2}
regions. Future observations to extend the MAGPIS coverage area will provide
the basis for a comprehensive view of massive star birth and death in the 
Milky Way.

\acknowledgments

DJH and RHB acknowledge the support of the National Science Foundation
under grants AST-05-07598 and AST-02-6-55; DJH was also supported
in this work by NASA grant NAG5-13062.  RHB's work was supported
in part under the auspices of the US Department of Energy by Lawrence
Livermore National Laboratory under contract W-7405-ENG-48.  RLW
acknowledges the support of the Space Telescope Science Institute,
which is operated by the Association of Universities for Research
in Astronomy under NASA contract NAS5-26555.

\clearpage
\LongTables

\begin{deluxetable}{rccrccrrr}
\tablecolumns{9}
\tablewidth{0pc}
\tabletypesize{\scriptsize}
\tablecaption{Catalog of Discrete Sources}
\tablehead{
\colhead{Name ($l+b$)} & \colhead{$S_p$(20cm)\tablenotemark{a}} & \colhead{$S_i$(20cm)\tablenotemark{b}} & 
\colhead{RMS} & \colhead{Major\tablenotemark{c}} & \colhead{Minor\tablenotemark{c}} & 
\colhead{$\theta$} & \colhead{S(8$\mu$m)\tablenotemark{d}} & \colhead{S(21$\mu$m)\tablenotemark{d}}\\
\colhead{\arcdeg} & \colhead{mJy/beam} & \colhead{mJy} & 
\colhead{mJy} & \colhead{\arcsec} & \colhead{\arcsec} & 
\colhead{\arcdeg} & \colhead{Jy} & \colhead{Jy} \\
\colhead{(1)} & \colhead{(2)} & \colhead{(3)} & 
\colhead{(4)} & \colhead{(5)} & \colhead{(6)} & 
\colhead{(7)} & \colhead{(8)} & \colhead{(9)}
}
\startdata
10.74968${-}$0.41783 & \phn1.69 & \phn\phn7.07\phm{E} &    0.291 & 15.1 & \phn9.2 &   54.2 &  \nodata  &  \nodata  \\
10.78586${-}$0.14376 & 12.86 & \phn15.14\phm{E} &    0.224 & \phn6.7 & \phn5.9 &   36.7 &  \nodata  &  \nodata  \\
10.80070${+}$0.15285 & \phn5.28 & \phn\phn6.51\phm{E} &    0.162 & \phn6.8 & \phn6.1 &   18.0 &  \nodata  &  \nodata  \\
10.81109${+}$0.75435 & \phn1.92 & \phn\phn2.49\phm{E} &    0.293 & \phn8.5 & \phn5.1 &   83.2 &  \nodata  &  \nodata  \\
10.82677${-}$0.01064 & \phn3.98 & \phn26.58E &    0.275 & 17.5 & 12.8 &   16.3 & $   1.10$ & $   1.56$ \\
10.82960${-}$0.01145 & \phn2.05 & \phn\phn6.68E &    0.275 & 15.6 & \phn7.0 &   48.5 &  \nodata  &  \nodata  \\
10.85201${-}$0.44069 & \phn6.61 & \phn\phn9.99\phm{E} &    0.222 & \phn7.3 & \phn6.9 &  116.2 &  \nodata  &  \nodata  \\
10.86448${-}$0.32989 & \phn2.32 & \phn\phn3.28\phm{E} &    0.365 & \phn7.3 & \phn6.5 &   60.3 &  \nodata  &  \nodata  \\
\enddata
\tablecomments{Table 2 is published in its entirety in the electronic edition of the {\it Astronomical
Journal} and is also available on the MAGPIS website (\url{http://third.ucllnl.org/gps}).  A portion is
shown here for guidance regarding its form and content.}
\tablenotetext{a}{Peak flux density at 20~cm.}
\tablenotetext{b}{Integrated flux density at 20~cm. An appended `E' indicates that
the Gaussian model is inadequate and that the value is probably inaccurate.}
\tablenotetext{c}{Full-width at half-maximum for major and minor axes
of the elliptical Gaussian fit. The CLEAN beam ($6\farcs2\times5\farcs4$)
determines the size for unresolved sources.}
\tablenotetext{d}{The 8$\mu$m and 21$\mu$m infrared fluxes come from the MSX6C catalog
(Egan et al. 2003).}
\end{deluxetable}

\clearpage

\begin{deluxetable}{rcccrrcl}
\tablecolumns{8}
\tablewidth{0pc}
\tabletypesize{\scriptsize}
\tablecaption{Catalog of Diffuse Emission Regions}
\tablehead{
\colhead{Name ($l+b$)} & \colhead{Box Size} & \colhead{Scale} & 
\colhead{$S_p$} & \colhead{$l_p$} & \colhead{$b_p$} & 
\colhead{$S_i$} & \colhead{Comment}\\
\colhead{\arcdeg} & \colhead{arcmin} & \colhead{mJy/beam} & 
\colhead{mJy/beam} & \colhead{\arcdeg} & \colhead{\arcdeg} & 
\colhead{Jy} & \\
\colhead{(1)} & \colhead{(2)} & \colhead{(3)} & 
\colhead{(4)} & \colhead{(5)} & \colhead{(6)} & 
\colhead{(7)} & \colhead{(8)}
}
\startdata
 5.1917${-}$0.2833 & \phn2\phd\phn & \phn15\phd\phn & \phn17.0 &   5.1903 & ${-}$0.2838 & \phn\phn1.05\phn &  \\
 5.4444${+}$0.2056 & \phn1\phd\phn & \phn\phn3\phd\phn & \phn\phn2.3 &   5.4466 & ${+}$0.2052 & \phn\phn0.136 &  \\
 5.4667${+}$0.3833 & \phn6\phd\phn & \phn\phn1\phd\phn & \phn\phn1.7 &   5.5155 & ${+}$0.3474 & \phn\phn1.92\phn &  \\
 5.4750${-}$0.2458 & \phn6\phd\phn & \phn\phn7\phd\phn & 114.3 &   5.4749 & ${-}$0.2432 & \phn\phn4.04\phn &  \\
 5.5250${+}$0.0292 & \phn4\phd\phn & \phn\phn5\phd\phn & \phn\phn7.5 &   5.5228 & ${+}$0.0285 & \phn\phn1.04\phn &  \\
 5.9000${-}$0.4333 & 20\phd\phn & \phn20\phd\phn & 199.5 &   5.8851 & ${-}$0.3924 & \phn56.9\phn\phn &  \\
 5.9306${-}$0.0972 & \phn2\phd\phn & \phn\phn4\phd\phn & \phn\phn3.8 &   5.9301 & ${-}$0.0992 & \phn\phn0.481 &  \\
 5.9500${+}$0.0833 & 25\phd\phn & \phn10\phd\phn & 175.7 &   5.9440 & ${+}$0.2013 & \phn47.9\phn\phn &  \\
\enddata
\end{deluxetable}

\clearpage

\begin{deluxetable}{rcccc}
\tablecolumns{5}
\tablewidth{0pc}
\tabletypesize{\scriptsize}
\tablecaption{High-Probability Supernova Remnant Candidates}
\tablehead{
\colhead{Name ($l+b$)} & \colhead{Diameter} & \colhead{Box Size} & \colhead{$S_p$} & \colhead{$S_i$} \\
\colhead{\arcdeg} & \colhead{arcmin} & \colhead{arcmin} & \colhead{mJy/beam} & \colhead{Jy} \\
\colhead{(1)} & \colhead{(2)} & \colhead{(3)} & \colhead{(4)} & \colhead{(5)}
}
\startdata
 6.4500${-}$0.5583 & \phn3.3 & \phn6.0 & \phn\phn4.7 & \phn6.64\phn \\
 6.5375${-}$0.6028 & \phn5.0 & \phn9.0 & \phn16.9 & \phn9.42\phn \\
 7.2167${+}$0.1833 & \phn6.5 & 10.0 & \phn17.2 & \phn9.10\phn \\
 8.3083${-}$0.0861 & \phn3.0 & \phn6.0 & 263.5 & \phn7.00\phn \\
 8.8583${-}$0.2583 & \phn4.0 & \phn6.0 & \phn\phn3.4 & \phn4.79\phn \\
 9.6833${-}$0.0667 & \phn8.5 & 12.0 & \phn\phn7.2 & 12.5\phn\phn \\
10.8750${+}$0.0875 & \phn2.8 & \phn6.0 & \phn\phn4.9 & \phn4.05\phn \\
11.1639${-}$0.7167 & \phn7.0 & 10.0 & \phn\phn1.9 & \phn3.11\phn \\
\enddata
\end{deluxetable}

\end{document}